\documentclass[10pt,conference]{IEEEtran}
\usepackage[font=footnotesize]{caption}
\usepackage{graphicx,times,cite,amsmath, amssymb, epsfig}
\usepackage{amsmath}
\usepackage{amsfonts}
\usepackage{color}
\usepackage{subfig}
\usepackage{float}
\usepackage[noend]{algorithmic}
\usepackage{algorithm}
\newtheorem{remark}{\underline{Remark}}
\newtheorem{lemma}{Lemma}

\usepackage{color}
\usepackage{subfig}
\usepackage{float}

\newlength{\figwidth}
\IEEEoverridecommandlockouts 
\begin{document}
\title{
Performance Analysis for Multi-User  Holographic MIMO Downlink with Matched Filter Precoding}
\author{
\IEEEauthorblockN{
Gayathri Shekar\IEEEauthorrefmark{1}, 
Saman Atapattu\IEEEauthorrefmark{1},
Prathapasinghe Dharmawansa\IEEEauthorrefmark{2}, and
Kandeepan Sithamparanathan\IEEEauthorrefmark{1} 
}
\IEEEauthorblockA{
\IEEEauthorrefmark{1}Department of Electrical and Electronic Engineering, School of Engineering, RMIT University, Melbourne,  Australia\\
\IEEEauthorrefmark{2}Centre for Wireless Communications, University of Oulu, Finland.\\
Email: 
\IEEEauthorrefmark{1}\{gayathri.shekar,\,saman.atapattu,\,kandeepan.sithamparanathan\}@rmit.edu.au,  
\IEEEauthorrefmark{2}pkaluwad24@univ.yo.oulu.fi. 
}
}

\maketitle
\begin{abstract}
Holographic MIMO (HMIMO) has emerged as a promising solution for future wireless systems by enabling ultra-dense, spatially continuous antenna deployments. While prior studies have primarily focused on electromagnetic (EM) modeling or simulation-based performance analysis, a rigorous communication-theoretic framework remains largely unexplored. This paper presents the first analytical performance study of a multi-user HMIMO downlink system with matched filter (MF) precoding-a low-complexity baseline scheme. By incorporating multipath propagation, mutual coupling, and element excitation, we derive a novel closed-form expression for the MF signal-to-interference-plus-noise ratio (SINR) using an equivalent random variable model. Leveraging bivariate gamma distributions, we then develop tractable throughput approximations under full, partial, and no channel state information (CSI) scenarios. Additionally, we formulate a max–min beamforming problem to benchmark optimal user fairness performance. Numerical results validate the accuracy of the proposed framework and reveal that MF precoding achieves competitive performance with strong robustness to low SINR and CSI uncertainty.
\end{abstract}
\section{Introduction} \label{s:intro} 
In response to the growing demand for ubiquitous high-rate connectivity \cite{yang20196g}, massive MIMO has become a key technology for next-generation networks \cite{pan2017many}. However, the physical size of antenna arrays limits its scalability. Recent advances in metamaterials, metasurfaces, and antenna technologies-particularly the development of reconfigurable intelligent surfaces (RIS) \cite{10843375} and  large-aperture architectures like holographic MIMO (HMIMO) has emerged as a compact alternative, enabling ultra-dense arrays that approximate continuous EM apertures \cite{huang2020holographic}. Although initial system-level studies confirm its feasibility \cite{gong2023holographic}, rigorous theoretical performance analysis remains limited, which is the focus of this paper.
Accurate channel modeling remains a key challenge in HMIMO due to the continuous nature of the antenna surface and the high density of elements \cite{an2023tutorial}. Traditional models based on simplified assumptions are not well suited for HMIMO, particularly at higher frequencies such as mmWave and THz \cite{ pizzo2022spatial, 9906802}. In such systems, closely spaced elements can interact strongly with each other, a phenomenon known as {\it mutual coupling}, which significantly affects the channel behavior and complicates performance analysis \cite{ 9300189}. 

To improve modeling accuracy, it is essential to consider the impact of mutual coupling and the physical characteristics of wave propagation. Electromagnetic Information Theory (EIT) provides a suitable framework by combining EM and information-theoretic  tools \cite{ 10500751}. 
In multi-user HMIMO (MU-HMIMO) systems, {\it precoding} design is critical to achieving high spectral efficiency and managing inter-user interference. 
In \cite{deng2022hdma}, a novel multiple access method demonstrated that ZF precoding can asymptotically approach channel capacity. A distance-aware hybrid precoding architecture was proposed in \cite{9860745}, exploiting spatial characteristics of user locations. Beyond algorithmic design, HMIMO systems require joint optimization of multiple physical and signal processing parameters, including antenna excitation, element placement, mutual coupling, and precoding vectors. Recent studies such as \cite{9300189, 10893696} have addressed subsets of these parameters using EM-based modeling approaches.  

Although prior studies have examined precoding strategies such as Zero Forcing (ZF), Regularized ZF (RZF), and MMSE in HMIMO systems, much of this work is rooted in EM theory, emphasizing array excitation, element placement, and mutual coupling \cite{deng2022hdma}--\cite{10893696}. While some system-level studies combine EM and communication parameters, they often rely on numerical optimization and lack a unified analytical framework for performance evaluation \cite{9300189}. A rigorous treatment that jointly captures EM characteristics (e.g., coupling, excitation) and communication-theoretic aspects (e.g., channel randomness, beamforming, noise) remains largely unexplored.
 {\it This paper addresses this gap by providing an analytical performance study of HMIMO downlink systems under matched filter (MF) precoding, a widely used low-complexity baseline, that has not been previously examined in the HMIMO context}.
The main contributions are:
i) A novel MF SINR expression is derived by incorporating multipath propagation, mutual coupling, and excitation effects, represented via an equivalent random variable;
ii) Closed-form throughput approximations are developed using bivariate gamma distributions to capture coupling-induced channel correlation under full, partial, and no CSI;
iii) A max–min fairness problem is formulated to obtain an optimal beamforming benchmark for user fairness, which has not been explored in prior HMIMO studies; and
iv) Numerical results validate the analysis, demonstrating MF precoding’s robustness at low SINR and under CSI imperfections, and quantify its tradeoff against the optimal solution.

\begin{figure}[t]
    \centering
\includegraphics[width=1.2\figwidth]{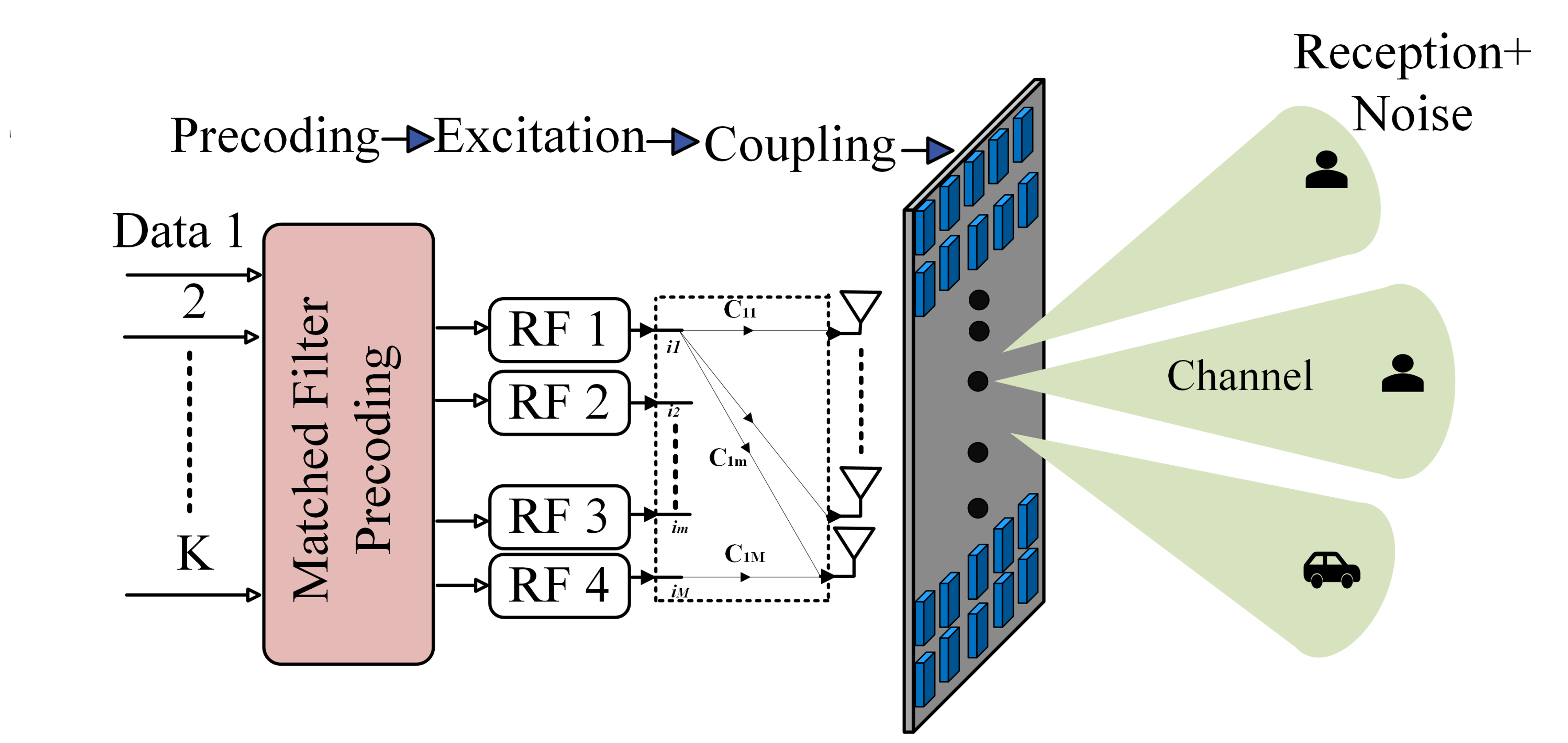}
    \caption{A multi-user holographic MIMO (MU-HMIMO) system: Precoding → Excitation → Coupling → Channel → Reception + Noise.}\label{fig 1-System Model}
\end{figure}

\section{System Model}
\subsection{Network Model}\label{ss_network}
We consider a MU-HMIMO system, as depicted in Fig.~\ref{fig 1-System Model}, where a base station (BS) equipped with \( M \) antenna elements serves \( K \) single-antenna users. The transmitted signal vector is \( \mathbf{x} = \{x_k\}_{k=1}^K \in \mathbb{C}^{K \times 1} \), where each \( x_k \) is an independent, energy-normalized  symbol, i.e., \( \mathbb{E}[|x_k|^2] = 1 \). The BS applies a precoding to the transmit signal. The antenna current excitation vector is denoted by \( \mathbf{i} \in \mathbb{C}^{M \times 1} \), and the precoded signal is given by \( \mathbf{x}_e = \mathbf{I} \mathbf{W} \mathbf{x} \in \mathbb{C}^{M \times 1} \), where \( \mathbf{W} \in \mathbb{C}^{M \times K} \) is the precoding matrix, and \( \mathbf{I} = \text{diag}(\mathbf{i}) \in \mathbb{C}^{M \times M} \) is the diagonal matrix representing the antenna excitation currents. 

Additionally, mutual coupling among the antenna elements introduces signal distortion, which modifies the transmitted signal at the HMIMO array. Specifically, the transmitted signal after the coupling effects is expressed as \( \mathbf{x}_c = \mathbf{C} \mathbf{x}_e \in \mathbb{C}^{M \times 1} \)~\cite{10628002}, where \( \mathbf{C} \in \mathbb{C}^{M \times M} \) is the coupling matrix. Elements of \( \mathbf{C} \) are modeled as \( [\mathbf{C}]_{n,m} = \mathrm{sinc} \left( {2\pi \|\mathbf{a}_n - \mathbf{a}_m\|}/{\lambda} \right) \), where \( \lambda \) is the wavelength, and \( \|\mathbf{a}_n - \mathbf{a}_m\| \) is the distance between the \( n \)th and \( m \)th antenna elements~\cite{9300189}.

\subsection{Signal Model for Matched Filter (MF) Precoding}
The  multipath fading channel between the HMIMO BS and User \( k \) is represented by \( \mathbf{\alpha}_k = \{\alpha_{k,m}\}_{m=1}^M \in \mathbb{C}^{1 \times M} \). The full channel matrix for all \( K \) users is then \( \mathbf{A} = [\mathbf{\alpha}_1, \dots, \mathbf{\alpha}_K]^T \in \mathbb{C}^{K \times M} \). The received signal vector for all users is given by $\mathbf{y}=\mathbf{A} \mathbf{x}_e+ \mathbf{n}\in \mathbb{C}^{K \times 1}$, which can be rewritten as  
\begin{align}
\mathbf{y} = \left(\mathbf{A} \mathbf{C} \mathbf{I}\right)\mathbf{W} \mathbf{x}+ \mathbf{n} 
=  \mathbf{H}\mathbf{W} \mathbf{x}+ \mathbf{n},
\label{eq_received_signal}
\end{align}
where \( \mathbf{n} \) is additive white Gaussian noise (AWGN), modeled as  independent and identically distributed (i.i.d.) circularly symmetric complex Gaussian RVs (RVs) across users with zero mean $N_0$ variance, \( \mathbf{n} \in \mathbb{C}^{K \times 1} \sim \mathcal{CN}(\mathbf{0}, N_0 \mathbf{I}_K) \). 
The effective end-to-end channel  is \( \mathbf{H} = \mathbf{A} \mathbf{C} \mathbf{I} \in \mathbb{C}^{K \times M} \),  where each row \( \mathbf{h}_k = \mathbf{\alpha}_k \mathbf{CI} \in \mathbb{C}^{1 \times M} \) represents the User \( k \) channel. 

We adopt MF precoding for its low complexity, signal power gain in low-SNR regimes, and analytical tractability, enabling clear performance characterization and serving as a baseline for advanced precoders (ZF, RZF, MMSE). 
For MF, each precoding vector \( \mathbf{w}_k \) is normalized to satisfy the total power constraint. Specifically, the precoder for User \( k \) is given by \( \mathbf{w}_k = (\sqrt{P/M}) \mathbf{h}_k^H \), where \( \mathbf{h}_k \) is the channel vector for User \( k \). The overall precoding matrix is then \( \mathbf{W} = (\sqrt{P/M}) \mathbf{H}^H \in \mathbb{C}^{M \times K} \). Accordingly, the received signal at User \( k \) can be expressed as \cite{atapattu2017exact} 
\begin{align}
y_k 
= \sqrt{{P}/{M}} {\bf h}_k {\bf h}_k^H x_k 
+\sqrt{{P}/{M}}  \sum_{\substack{j=1}}^{K-1} {\bf h}_k {\bf h}_j^H x_j 
+ n_k,
\label{eq:yk}
\end{align}
where \( n_k \sim \mathcal{CN}(0, N_0) \) is the AWGN noise term for User \( k \).

\subsection{Equivalent SINR Models}
For User \( k \), we assume the channels in \( \mathbf{\alpha}_k \) are i.i.d. and \( \mathbf{\alpha}_k \sim \mathcal{CN}(\mathbf{0}, \sigma_k^2 \mathbf{I}_M) \), where \( \sigma_k^2 \) is the variance for User \( k \). The variances \( \sigma_k^2 \) may vary across users, i.e., \(\sigma_k\neq\sigma_j\) for User~$k$ and User~$j$. These variances reflect different large-scale fading effects, e.g., distance-dependent path loss between users and the HMIMO BS. Then, SINR of User \( k \) can be given as 
\begin{align}\label{eq_sinr_org}
\gamma_k &\stackrel{}{=} \frac{\rho \left|{\bf h}_k {\bf h}_k^H\right|^2}{1 + \rho \sum_{\substack{j=1 \\ j \ne k}}^{K} \left|{\bf h}_k {\bf h}_j^H\right|^2}, \text{ where }\rho=\frac{P}{MN_0}.
\end{align}
Here $\rho$ can be interpreted  as the power-normalized factor.  
To facilitate tractable analysis, we separate the deterministic component of the HMIMO channel  correlation matrix as
\begin{equation}
    \mathbf{Q} = \mathbf{R}\mathbf{R}^{H},  \text{where}\quad \mathbf{R} = \mathbf{C}\mathbf{I}.
    \label{correlationmatrix}
\end{equation}
 Since \( \mathbf{h}_k = \mathbf{\alpha}_k \mathbf{CI} = \mathbf{\alpha}_k\mathbf{R}\), it follows that \( \mathbf{h}_k \mathbf{h}_k^H = \mathbf{\alpha}_k \mathbf{Q} \mathbf{\alpha}_k^H  \),   Let \( q_{m,n} \) denote the \( (m,n) \)-th entry of \( \mathbf{Q} \). Similarly, the cross-term between users \( k \) and \( j \) is \( \mathbf{h}_k \mathbf{h}_j^H = \mathbf{\alpha}_k \mathbf{Q} \mathbf{\alpha}_j^H \). The correlation introduced by \( \mathbf{Q} \), along with the presence of \( \mathbf{h}_k \) in both the numerator and denominator of \eqref{eq_sinr_org}, complicates direct analysis. To improve tractability, we reformulate the SINR expression into an analytically equivalent form, following the approach in \cite{atapattu2017exact} originally developed for uncorrelated cases.
For analytical tractability, we model the cross-term \( \mathbf{h}_k \mathbf{h}_j^H \) as
$\mathbf{h}_k \mathbf{h}_j^H \sim \mathbf{\alpha}_k \mathbf{Q}^2 \mathbf{\alpha}_k^H\sum_{j=1}^{K-1} \sigma_j^2 |y_j|^2$
where \( y_j \sim \mathcal{CN}(0,1) \) for \( j = 1, \dots, K-1 \), capturing the randomness and reflecting the independent nature of the interference terms. This arises from the fact that \( \mathbf{\alpha}_k \mathbf{Q}  \mathbf{\alpha}_j^H \) follows the distribution \( \mathbf{\alpha}_k \sim \mathcal{CN}(\mathbf{0}, \sigma_k^2 \mathbf{I}_M) \),  which implies the conditional distribution
$\mathbf{\alpha}_k \mathbf{Q}  \mathbf{\alpha}_j^H \mid \mathbf{\alpha}_k \sim \mathcal{CN}(\mathbf{0}, \sigma_j^2 \mathbf{\alpha}_k \mathbf{Q}^2 \mathbf{\alpha}_k^H)$. 

This model facilitates efficient analysis of the interference terms while explicitly capturing the impact of mutual coupling through the correlation matrix $\mathbf{Q}$.  Accordingly, the SINR  $\gamma_k$ can be equivalently represented in distribution as
\begin{align}\label{eq_sinr_eqi}
\Tilde{{\gamma}}_k 
&\stackrel{}{:=} \frac{\rho|\mathbf{\alpha}_k \mathbf{Q} \mathbf{\alpha}_k^H|^2}{1 +\rho (\mathbf{\alpha}_k \mathbf{Q}^2 \mathbf{\alpha}_k^H) \sum_{j=1}^{K-1} \sigma_j^2 |y_j|^2},
\end{align}
where  the quadratic forms are given by,
\[
\mathbf{\alpha}_k \mathbf{Q} \mathbf{\alpha}_k^H = \sum_{i=1}^M \lambda_i |\tilde{\alpha}_{k,i}|^2 \text{ and }
\mathbf{\alpha}_k \mathbf{Q}^2 \mathbf{\alpha}_k^H = \sum_{i=1}^M \lambda_i^2 |\tilde{\alpha}_{k,i}|^2.
\]
To enable tractable analysis, we perform eigenvalue decomposition (EVD) of the correlation matrix as \( \mathbf{Q} = \mathbf{U} \mathbf{\Lambda} \mathbf{U}^H \), where \( \mathbf{U} \in \mathbb{C}^{M \times M} \) is unitary and \( \mathbf{\Lambda} = \mathrm{diag}(\lambda_1, \dots, \lambda_M) \in \mathbb{R}^{M \times M} \) contains eigenvalues $\lambda_i$s. Defining the rotated channel vector \( \tilde{\mathbf{\alpha}}_k := \mathbf{\alpha}_k \mathbf{U} \), and thus we have the quadratic forms mentioned above.

Since \( \mathbf{U} \) is unitary and \( \mathbf{\alpha}_k \sim \mathcal{CN}(\mathbf{0}, \sigma_k^2 \mathbf{I}) \), the projected components \( \tilde{\alpha}_{k,i} \) remain i.i.d. as \( \mathcal{CN}(0, \sigma_k^2) \), preserving the distribution. Substituting these into \eqref{eq_sinr_eqi}, $\Tilde{{\gamma}}_k$  becomes
\begin{align}
\Tilde{\gamma}_k &\!\!:=\! \frac{\rho \left( \sum_{i=1}^M \lambda_i |\tilde{\alpha}_{k,i}|^2\right)^2}{1 + \rho \sum_{i=1}^M \lambda_i^2 |\tilde{\alpha}_{k,i}|^2 \sum_{j=1}^{K-1} \sigma_j^2 |y_j|^2} 
\!=\! \frac{\rho X_1^2}{1 + \rho X_2 Y} \label{eq_sinr_eqi2}\\
X_1 &\!=\! \sum_{i=1}^M \lambda_i |\tilde{\alpha}_{k,i}|^2;\, X_2\!=\! \sum_{i=1}^M \lambda_i^2 |\tilde{\alpha}_{k,i}|^2;\,Y \!=\! \sum_{j=1}^{K-1} \sigma_j^2 |y_j|^2. \nonumber
\end{align}
\begin{remark}[Statistical Equivalence of SINR Representations]
Although the SINR \( \gamma_k \) in \eqref{eq_sinr_org} represents an instantaneous value, the SINR \( \Tilde{\gamma}_k \) in \eqref{eq_sinr_eqi} and \eqref{eq_sinr_eqi2} does not retain this instantaneous interpretation. Nevertheless, \( \gamma_k \) and \( \Tilde{\gamma}_k \) are statistically equivalent, i.e., they share the same distribution. This reformulation is thus justified as it enables a more tractable analysis of average performance metrics.
\end{remark}
\section{Throughput Analysis for MF Precoding}
Throughput is a fundamental metric in wireless communications. Leveraging the tractable SINR in \eqref{eq_sinr_eqi2}, we analyze the throughput of MF precoding in a MU-HMIMO system under three practical channel state information (CSI) scenarios: (i) full CSI, (ii) partial CSI, and (iii) no CSI at the BS.

\subsection{With Full CSI}\label{ss_mf_csi}
In the full CSI scenario, the HMIMO BS has perfect knowledge of the channel matrix  \( \mathbf{A} = [\mathbf{\alpha}_1, \dots, \mathbf{\alpha}_K]^T \in \mathbb{C}^{K \times M} \). 
For User~$k$, the average throughput can be expressed as  
    $\bar{R}_k = \mathbb{E}_{\mathbf{A}}\left[\ln\left(1 + \gamma_k\right)\right]$
where $\gamma_k$ is the instantaneous SINR in \eqref{eq:yk}. Using \eqref{eq_sinr_eqi2}, this can be equivalently expressed as   
\begin{align}\label{eq_avg_rate_k1}
    \bar{R}_k = \mathbb{E}_{X_1,X_2,Y}\left[\ln\left(1 + \frac{\rho  X_1^2}{1 + {\rho } X_2 Y}\right)\right]. 
\end{align}
An exact evaluation of \( \bar{R}_k \) is challenging due to the correlation between the RVs \( X_1 \) and \( X_2 \), the multiplicative interaction with the additional variable \( Y \), and their presence within a nonlinear logarithmic function. To address this complexity, we adopt an analytical approximation strategy:
\begin{itemize}
    \item Since \( X_1 \), \( X_2 \), and \( Y \) are each sums of weighted exponential RVs, they can be reasonably approximated by gamma distributions through moment matching \cite{oguntunde2014sum}. For tractability, we define \( X_3 = X_2 Y \), and approximate both \( X_1 \) and \( X_3 \) as gamma-distributed RVs. Due to the correlation between \( X_1 \) and \( X_2 \), the variables \( X_1 \) and \( X_3 \) are also correlated. Thus, their joint distribution is modeled using a bivariate gamma distribution \cite{6059452}.    
    \item To handle the nonlinearity due to $\log(\cdot)$, we use Jensen’s inequality to derive an approximation for \( \bar{R}_k \) as  
    \begin{align}\label{eq_apx_avg_rate_k1}
        \bar{R}_k \approx \ln\left(1 + \mathbb{E}_{X_1,X_3}\left[\frac{\rho X_1^2}{1 + \rho X_3 }\right]\right),
    \end{align}
    Jensen’s inequality is a widely used technique, especially effective in MIMO systems, where the law of large numbers and averaging over many RVs lead to tight approximations.
\end{itemize}
Using the analytical procedure outlined above, Lemma~\ref{Lemma_avg_Rk} provides an approximation for $\bar{R}_k$.

\begin{lemma}[Average User Throughput with MF Precoding] \label{Lemma_avg_Rk}
Consider a MU-HMIMO system employing MF precoding at a BS with \( M \) antennas serving \( K \) users. Under the effect of mutual coupling, the average throughput of User~\( k \), for \( k = 1, \dots, K \), can be approximated as
\begin{align}\label{eq_apx_avg_rate_k}
   \bar{R}_k &\approx \ln\left(1 + \Sigma\right), \quad \text{ where }\\
\Sigma & =\frac{\theta^2(1-\eta)^2}{\rho^{\mu+1}\phi^\mu} \sum_{i=0}^\infty \frac{}{} 
\sum_{j=0}^\infty \frac{(\nu)_i \eta^i(\mu - \nu)_j \eta^j}{i!(\mu + i)_j\left(\rho(1-\eta)\phi\right)^{i+j}} 
 \nonumber\\
&\qquad\quad\times
\frac{\Gamma(\nu + i + 2) \Gamma(\mu + i + j)}{ \Gamma(\nu + i) \Gamma(\mu + i)}e^{\frac{1}{{\rho} \phi(1 - \eta)}}  \nonumber\\
&\qquad\quad\times 
\Gamma\left(1 - (\mu + i + j), \frac{1}{{\rho}\phi(1-\eta)}\right) \nonumber\\
\nu & = \frac{L(\lambda_i,1)^2}{L(\lambda_i,2)},  \mu = \frac{(K-1)  L(\lambda_i,2)^2}{\sigma_k^4 (K-1) L(\lambda_i,2)^2 + K L(\lambda_i,4)}, \nonumber\\
    \theta &\! =\! \frac{\sigma_k^2L(\lambda_i,2)}{L(\lambda_i,1)}, 
\phi \!=\! \frac{\sigma_k^4 (K-1) L(\lambda_i,2)^2 + K L(\lambda_i,4)}{(K-1) L(\lambda_i,2)}, \nonumber\\
\eta & = 
\frac{\sigma_k^2\sqrt{L(\lambda_i,2) }  G(\sigma_j,2)}
{ 
\sqrt{L(\lambda_i,4) G(\sigma_j,2) + 2\sigma_k^2L(\lambda_i,4)  G(\sigma_j,4)}},\nonumber
\end{align}
with $L(\mathbf{\lambda},n) = \sum_{i=1}^{M} \lambda_i^n$ where $\mathbf{\lambda}=(\lambda_1,\cdots,\lambda_M)$ and $G(\sigma_j,n) = \sum_{j=1}^{K-1} \sigma_j^n$. 
\end{lemma}
\begin{IEEEproof}
To evaluate $\mathbb{E}_{X_1,X_3}[\cdot]$ in \eqref{eq_apx_avg_rate_k1}, we approximate the correlated random variables $X_1$ and $X_3$ using gamma distributions via moment matching~ \cite{oguntunde2014sum}, yielding $X_1 \sim \text{Gamma}(\nu, \theta)$ and $X_3 \sim \text{Gamma}(\mu, \phi)$. To capture their correlation with correlation coeffcient \(\eta\), the joint distribution $f_{X_1, X_3}(x_1, x_3)$ is modeled using a bivariate gamma distribution~\cite{6059452}, given by
\begin{equation*}
\begin{split}
& f(x_1, x_3) 
= \sum_{i=0}^{\infty} \frac{(\nu)_i \eta^i}{i!(1-\eta)^{-\mu}} 
\frac{\left(\phi (1 - \eta)\right)^{-(\mu + i)}}{\left( \theta (1 - \eta)\right)^{\nu + i}} \frac{x_1^{\nu+i-1}}{\Gamma(\nu+i)}   \\
&\times 
\frac{x_3^{\mu+i-1}}{\Gamma(\mu+i)}  e^{-\frac{x_1}{\theta(1-\eta)} - \frac{\mu}{\phi(1-\eta)}} 
{}_1F_1\left[\mu-\nu, \mu+i, \frac{\eta x_3}{\phi(1-\eta)}\right].
\end{split}
\end{equation*}
We can now evaluate $\mathbb{E}_{X_1,X_3}[\cdot]$ as 
\begin{equation*}
\mathbb{E}\left[\frac{\rho X_1^2}{1 + \rho X_3}\right] = \int_{x_3} \int_{x_1} \frac{\rho x_1^2}{1 + \rho x_3} f_{X_1, X_3}(x_1, x_3) \, dx_1 \, dx_3
\label{eq:expected_value_Z}
\end{equation*}
To evaluate the expression, we first integrate with respect to \( x_1 \), which results in \( (\theta - \theta \eta)^{\nu + i + 2} \Gamma(\nu + i + 2) \). Next, by expanding the confluent hypergeometric function \( {}_1F_1(a; b; z) = \sum_{j=0}^{\infty} \frac{(a)_j z^j}{(b)_j j!} \), the second integral is expressed as \( I_2 = \rho \int_0^{\infty} {x_3^{i + j + 1}}e^{-x_3 / (\phi - \phi \eta)}/({x_3 + i})  dx_3 \). Finally, we obtain \eqref{eq_apx_avg_rate_k} referring \cite[Eq.~(10.3.383)]{gradshteyn2014table}.
\end{IEEEproof}

\begin{remark}[On the Analytical Approximation]
The expression originates from a rapidly converging double series of the bivariate Gamma distribution \cite{piboongungon2005bivariate}. By further approximating the involved RVs as Gamma distributed and applying Jensen's inequality, we obtain a tight estimate across a wide SNR range even for moderate $M$ and $K$, as validated in Section~\ref{s:sim}.
\end{remark}

As \( M \) and \( K \) increase, acquiring full CSI for all users becomes increasingly challenging due to overhead and complexity. In such large-scale systems, alternative strategies based on partial or no CSI become more suitable.

\subsection{With Partial CSI and No CSI}\label{ss_pcsi}

For the partial CSI scenario, we assume only second-order channel information is available. Specifically, each channel vector \( \mathbf{\alpha}_k \) is zero-mean with known variance \(\sigma_k^2\) for all \( K \). Under this assumption, the MF precoder from the full CSI case is approximated as $\mathbf{h}_k^H \rightarrow \sigma_k \left( \mathbf{1}_M \mathbf{R} \right)^H \in \mathbb{C}^{M \times 1}$,  
where \( \mathbf{1}_M \in \mathbb{C}^{1 \times M} \) is an all-ones row vector, and \( \mathbf{R} \in \mathbb{C}^{M \times M} \) models the mutual coupling matrix. 
In the no CSI case, we further assume that the variances \( \sigma_k^2 \) are also unknown. Consequently, the precoder simplifies to  $\mathbf{h}_k^H \rightarrow \left( \mathbf{1}_M \mathbf{R} \right)^H$. 

Then, the SINR of User~\( k \) under both cases is given by  
\begin{align}\label{eq_pcsi_inst_sinr_k}
    \gamma_k = 
\begin{cases}
   \displaystyle \frac{\rho\sigma_k^2  \left| \mathbf{\alpha}_k \mathbf{Q} \mathbf{1}_M^T \right|^2}{1 + \rho\left| \mathbf{\alpha}_k  \mathbf{Q} \mathbf{1}_M^T \right|^2 G(\sigma_j,2)}, & \text{Partial CSI} \\[10pt]
   \displaystyle \frac{\rho\left| \mathbf{\alpha}_k \mathbf{Q} \mathbf{1}_M^T \right|^2}{1 + \rho\left| \mathbf{\alpha}_k  \mathbf{Q} \mathbf{1}_M^T \right|^2}, & \text{No CSI}
\end{cases}
\end{align}
where $\mathbf{1}_M^T$ becomes an all-ones column vector now. 
We define  $X =\left|\mathbf{\alpha}_k\mathbf{Q}\mathbf{1}_M^H\right|^2= \left|\sum_{i=1}^M \sum_{j=1}^M \alpha_{i,j} q_{i,j}\right|^2$, which is a weighted sum of complex Gaussian RVs. The result is also a complex Gaussian RV $\sum_{i=1}^M \sum_{j=1}^M \alpha_{i,j} q_{i,j}\sim \mathcal{CN}(0, \sigma_k^2 \left| \sum_{i=1}^M \sum_{j=1}^M q_{i,j} \right|^2)$.The magnitude squared of a complex Gaussian RV follows an exponential distribution $X \sim \text{Exp}(\beta)$,
with $\beta = \sigma_k^2 \left| \sum_{i=1}^M \sum_{j=1}^M q_{i,j} \right|^{-2}$. 

The SINR of user \(k\) is   ${\gamma}_k$ can thus be given as 
\begin{align}\label{eq_pcsi_inst_eqii_sinr_k}
    {\gamma}_k = 
\begin{cases}
   \frac{\rho   \sigma_k^2 X}{1 + \rho  G(\sigma_j,2) X  }, & \text{Partial CSI } \\
  \frac{\rho   X}{1 + \rho  X }, & \text{No CSI}
\end{cases}
\end{align}
The following Lemma~\ref{Lemma2} gives average throughput expression for User~$k$ under both partial CSI and no CSI. 
\begin{lemma}[User Rate under Partial/No CSI] \label{Lemma2}
For a MU-HMIMO setup of a BS with \( M \) antennas and \( K \) users, the average throughput of User~$k$, for $k=1,\cdots,K$, under mutual coupling  can be expressed  for partial and no CSI as 
\begin{align}\label{eq_pcsi_inst_eqii_sinr_k}
    \bar{R}_k &\approx \ln\left(1 + \bar{\Sigma}\right), \quad\text{where }\\
\bar{\Sigma} &= 
\begin{cases}
\begin{aligned}
&\frac{\sigma_k^2}{G(\sigma_j,2)} + 
\frac{\sigma_k^2 \beta}{\rho G(\sigma_j,2)^2} 
e^{\left(\frac{\beta}{\rho G(\sigma_j,2)}\right)} \nonumber\\
& \times \text{Ei}\left(-\frac{\beta}{\rho G(\sigma_j,2)}\right),
\end{aligned}
& \text{Partial CSI} \\
\begin{aligned}
&1 + \frac{\beta}{\rho} 
e^{\left(\frac{\beta}{\rho}\right)}  
\text{Ei}\left(-\frac{\beta}{\rho}\right),
\end{aligned}
& \text{No CSI}
\end{cases}
\end{align}
\end{lemma}
\begin{IEEEproof}
We apply Jensen's inequality to derive an approximated expression for $\bar{R}_k$, here the RV X follows exponential distribution with PDF given by $ f_X(x) = \beta \exp({-\beta x })$. Substituting the PDF and factoring out the constant terms, we obtain 
\begin{align*}
    \bar{\Sigma} = \frac{\rho\sigma_k^2 \beta}{G(\sigma_j,2)} \int_0^\infty \frac{x e^{-\beta x }}{1 + \rho G(\sigma_j,2) x} \, dx. 
\end{align*}
Finally, the solution for the above integral is obtained using \cite[Eq.~(5.3.353)]{gradshteyn2014table}.
\end{IEEEproof}

We already analyzed MU-HMIMO with MF precoding as a low-complexity baseline, deriving closed-form SINR and throughput under various CSI assumptions. We now extend to optimal beamforming via an optimization framework.

\begin{figure*}
    \centering
    \subfloat[Throughput vs SNR for  $K=8$.\label{figtpm}]{%
        \includegraphics[width=0.37\textwidth]{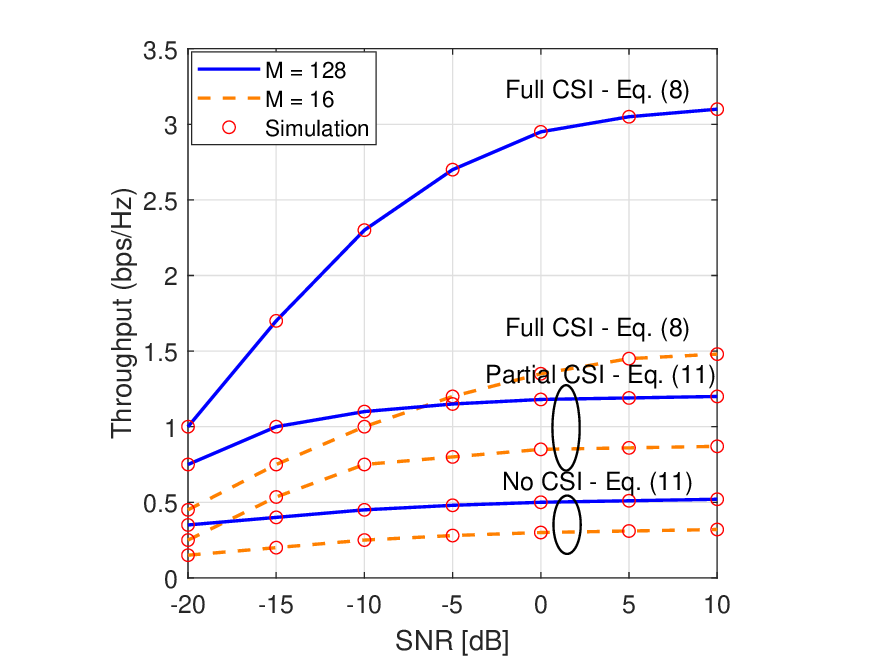}} 
      \hspace{-3.5em}
    \subfloat[Throughput vs  \(M\) for coupling effects. \label{fig_mctp}]{%
        \includegraphics[width=0.37\textwidth]{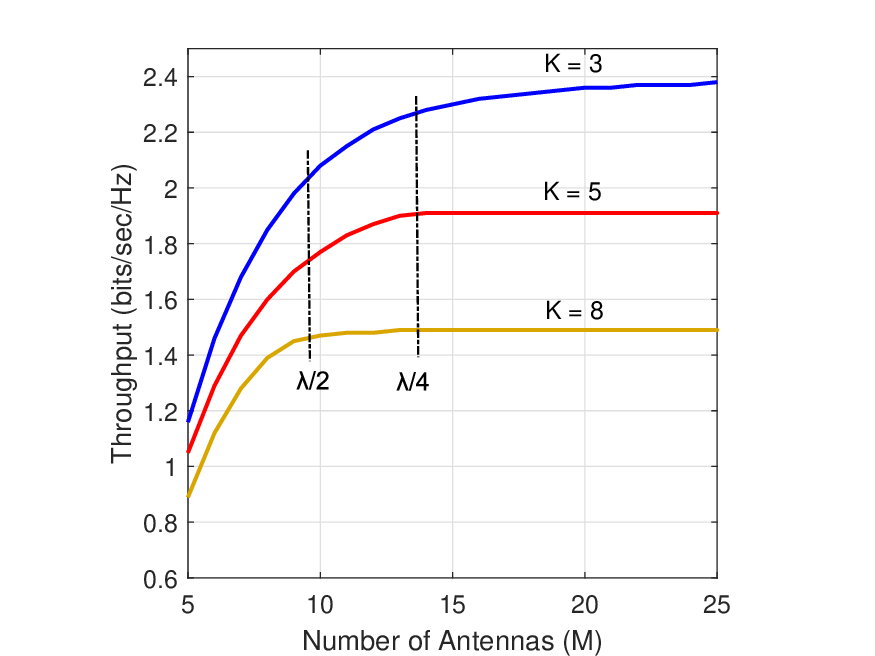}} 
         \hspace{-3.5em}
    \subfloat[Throughput vs \(M\) for different \(K\) values. \label{fig_tpfpncsi}]{%
        \includegraphics[width=0.37\textwidth]{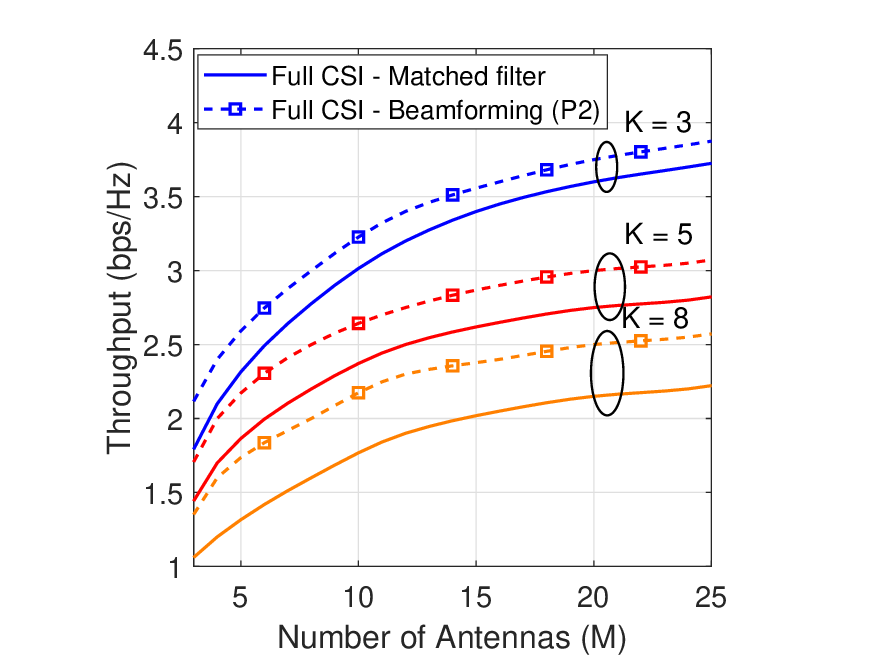}} 
    \caption{Average downlink MU-MIMO throughput under varying CSI, mutual coupling effects and precoding schemes.}
    \label{abcd}
\end{figure*}

\section{Beamforming Design via optimization}\label{opt}

Now, the precoder \( \mathbf{W} = [\mathbf{w}_1, \dots, \mathbf{w}_K] \in \mathbb{C}^{M \times K} \) in \eqref{eq_received_signal} is determined by formulating a user-fairness max-min SINR optimization problem. The goal is to maximize the minimum SINR across all users, subject to a total transmit power constraint. The SINR for User~\( k \) is now given by
\begin{align}
    \gamma_k = \frac{ \rho  \left| \mathbf{h}_k \mathbf{w}_k \right|^2 }
    { 1 + \rho  \sum\limits_{j=1}^{K-1} \left| \mathbf{h}_k \mathbf{w}_j \right|^2 },
\end{align}
where 
\( \mathbf{w}_k \in \mathbb{C}^{M \times 1} \) is the corresponding beamforming vector.

To this end, optimization problem is formulated as
\begin{align}
\text{(P1):} \quad & \underset{\{\mathbf{w}_k\}_{k=1}^K}{\text{maximize}} \quad \underset{k}{\min} \quad \gamma_k \\
& \text{subject to} \quad \sum_{k=1}^K \|\mathbf{w}_k\|_2^2 \leq P,
\end{align}
where \( P \) is the total transmit power budget. To express this in a more tractable form for optimization solvers, we can rewrite it using an epigraph variable \( t \) as
\begin{align}
\text{(P2):} \quad & \underset{\{\mathbf{w}_k\}_{k=1}^K, t}{\text{maximize}} \quad t \\
& \text{subject to} \quad \gamma_k \geq t, \quad \forall k \in \{1, \dots, K\}, \\
& \quad \quad \quad \quad \sum_{k=1}^K \|\mathbf{w}_k\|_2^2 \leq P. \label{eq:opt_reformulated}
\end{align}
This is a standard optimization problem with a linear objective in the epigraph variable \( t \). The SINR constraints can be reformulated as second-order cone (SOC) or semidefinite constraints, depending on the solution approach (e.g., via uplink-downlink duality or semidefinite programming (SDP) relaxations). In Section~\ref{s:sim}, we solve the problem using a bisection search algorithm combined with an SOCP reformulation of the SINR constraints, following the methods in \cite{10628002,boyd2004convex}.


\section{Numerical Results}\label{s:sim}
We evaluate the downlink performance of a multi-user HMIMO system at 1.6\,GHz. The transmit SNR is defined as $P/N_0$, with $P = 1$\,W and $N_0 = -104$\,dBm. Users are uniformly distributed in a $500$\,m-radius circular micro-cell. Distance-dependent path loss is modeled as $\sigma_k^2 = 1/d_k^\alpha$, where $d_k$ is the distance between the BS and user $k$, and $\alpha = 3.5$, thus we model fading as $\mathbf{\alpha}_k \sim \mathcal{CN}(\mathbf{0}, \sigma_k^2 \mathbf{I}_M)$. We model elements of antenna coupling  and excitation as  $[\mathbf{C}]_{n,m} = \mathrm{sinc} \left( 2\pi \|\mathbf{a}_n - \mathbf{a}_m\|/\lambda \right)$, following Sec.~~\ref{ss_network} and~~\cite{9300189}, and  $\mathbf{I} = \text{diag}(\mathbf{i})$ where $i_m = \exp(j\theta_m)$, with $\theta_m \sim \mathcal{U}[0,2\pi)$~\cite{10612761}. 

\subsection{Validation and performance comparison}
Fig.~\ref{figtpm} plots average throughput versus transmit SNR ($P/N_0$) for $M = 16$ and $M = 128$, with $K = 8$, under full, partial, and no CSI scenarios. The analytical expressions in \eqref{eq_apx_avg_rate_k} and \eqref{eq_pcsi_inst_eqii_sinr_k} closely match the simulated results across all SNR regimes, validating the accuracy of the proposed framework. Moreover, as the transmit SNR increases, the throughput saturates in all cases since both the desired signal and interference scale with transmit power. This reflects the inherent limitation of MF precoding, which lacks interference suppression capability. Further, CSI availability significantly impacts performance. At $10$\,dB SNR with $M=128$, the achieved throughputs are $3, 1.5$, and $1$\,bps/Hz for full, partial, and no CSI, respectively, indicating that full CSI yields $100$\% and $200$\% gains over partial and no CSI cases. 
\begin{figure}[!t]
    \centering
    \includegraphics[width=0.43\textwidth]{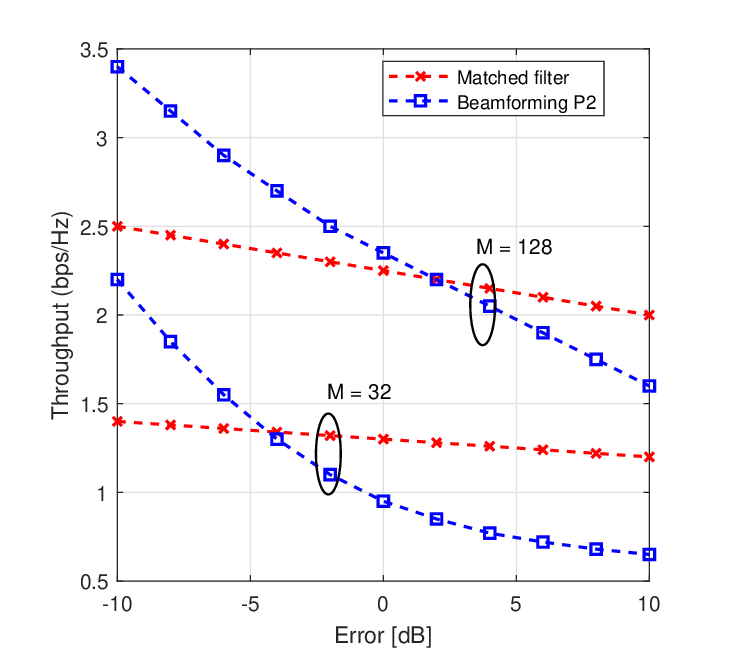}
    \caption{Throughput vs channel estimation error [dB].}
    \label{fig:tpvserror}
\end{figure}

\begin{figure}[!t]
    \centering
    \includegraphics[width=0.43\textwidth]{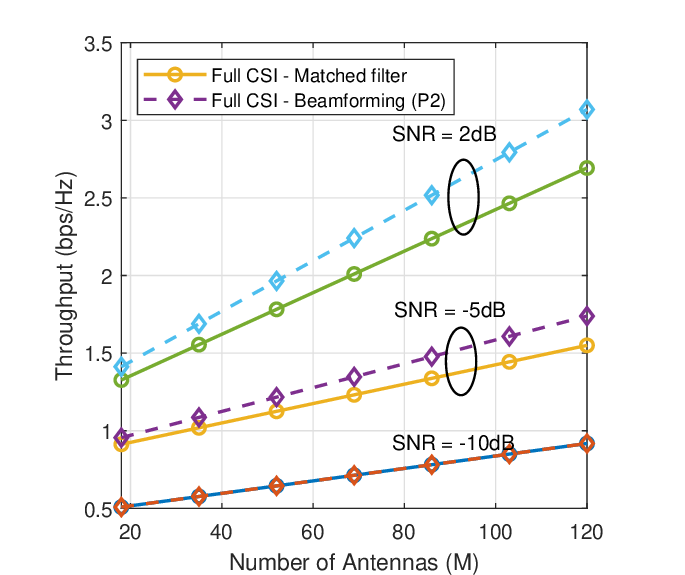}
    \caption{Throughput vs number of transmit antennas \(M\) for various SNR levels.}
    \label{fig:tpvsmvssnr}
\end{figure}
Fig.~\ref{fig_mctp} shows the average throughput versus the number of transmit antennas $M = 4, 9, 16, 25$, placed within a fixed $1\,\text{m}^2$ square aperture. Mutual coupling is modeled by $[\mathbf{C}]_{n,m} = \mathrm{sinc} \left( 2\pi \|\mathbf{a}_n - \mathbf{a}_m\|/\lambda \right)$. Throughput increases sharply as $M$ grows until the inter-element spacing approaches  half-wavelength $\lambda/2$, then the gain diminishes up to $\lambda/4$, and eventually saturates due to strong mutual coupling and high spatial correlation, limiting further benefits of array and diversity gains.

To compare MF precoding with max-min fairness optimal beamforming (BF) in Sec.~\ref{opt}, Fig.~\ref{fig_tpfpncsi} shows the average throughput versus the number of transmit antennas $M$ with constant element spacing for $K = 3, 5, 8$. Optimal BF consistently outperforms MF, with the performance gap widening as $K$ increases due to BF’s ability to better manage interference via power control. However, this comes at the cost of higher implementation complexity. For both schemes, throughput increases with the number of antennas $M$, owing to improved array gain and spatial diversity.

 \subsection{Impact of Imperfect CSI on Precoding}
 To assess the robustness of precoding schemes under imperfect CSI, we adopt the error model from \cite{8502064}, where the estimated channel is modeled as \(\mathbf{\hat{H}} = \mathbf{H} + \mathbf{E}\), where \( \mathbf{H} \in \mathbb{C}^{K \times M} \) denotes the true channel, and \( \mathbf{E} \sim \mathcal{CN}(\mathbf{0}, \sigma_e^2 \mathbf I) \in \mathbb{C}^{K \times M}\) represents the additive estimation error.  
  Precoding is then performed using \( \mathbf{\hat{H}} \) instead of the true channel \( \mathbf{H} \). 
Fig.~\ref{fig:tpvserror} shows the throughput versus channel estimation error (in dB) for $M = 32$ and $M = 128$. The optimization-based max-min precoding outperforms MF precoding under accurate CSI due to its interference-aware power allocation. However, its performance degrades sharply with increasing estimation error, as it relies heavily on precise CSI to maintain fairness. In contrast, MF precoding exhibits greater robustness under high estimation errors, since it only requires approximate channel direction information and avoids aggressive nulling. The crossover point between the two schemes shifts rightward with increasing $M$, reflecting improved resilience to estimation errors in larger arrays due to increased beamforming redundancy and array gain.
 
Fig.~\ref{fig:tpvsmvssnr} plots the average throughput versus the number of antennas $M$ for three received average SNR levels: $-10\,\mathrm{dB}$, $-5\,\mathrm{dB}$, and $2\,\mathrm{dB}$. At low SNR (e.g., $-10\,\mathrm{dB}$), MF and optimization-based precoding perform similarly across all $M$, making MF a practical, low-complexity choice in noise-limited regimes. As SNR increases, the optimization-based precoder achieves higher throughput by better suppressing multi-user interference and leveraging spatial diversity. Nonetheless, in realistic scenarios where received SNRs typically hover around $-5\,\mathrm{dB}$, MF remains an attractive solution due to its simplicity and robust performance.
 
\section{Conclusion}\label{con}
This work developed a rigorous analytical framework for evaluating the performance of multi-user HMIMO downlink systems under practical propagation conditions with MF precoding. A hybrid channel model was introduced to incorporate both deterministic effects, such as mutual coupling and excitation, and statistical variations due to multipath. A tractable SINR expression was derived, enabling closed-form throughput approximations under full, partial, and no CSI scenarios. To benchmark the MF baseline, an optimization-based beamforming scheme was formulated to maximize the minimum user SINR. Additionally, a robustness analysis under channel estimation errors demonstrated that MF precoding offers enhanced resilience compared to the optimal design. These results provide design insights for low-complexity, robust precoding strategies for next-generation MIMO systems.



\end{document}